\documentclass[aps,preprint,groupedaddress]{revtex4-1}
\usepackage{graphicx}
\usepackage{subfig}
\begin{document}


\title{Nonaxisymmetric instabilities in turbulent electromagnetic pumps}

\author{Paola Rodriguez-Imazio}

\affiliation{
Servicio Meteorológico Nacional, CONICET-SMN, Dorrego 4019, CABA 1425, 
Buenos Aires, Argentina}
\author{Christophe Gissinger}
\affiliation{Laboratoire de Physique (LPENS), Ecole Normale Superieure, CNRS, 24 rue Lhomond,75005 Paris, France}

\date{\today}

\begin{abstract}
The stability of  annular electromagnetic pumps (EMPs) is investigated through 3D direct numerical simulations. When induction effects dominate dissipative processes, linear induction EMPs become unstable and lead to non-axisymmetric states relatively different from what was predicted by previous theoretical models. We show that the 3D destabilization of the flow is deeply connected to the axisymmetric state, and always appears as a secondary bifurcation from a locally stalled flow, even if the applied magnetic field is axisymmetric. Finally, we model a configuration aiming to optimize the efficiency of the pump, by imposing electrical currents on both sides (inner and outer cylinders) of the pump. This  configuration increases the efficiency of the pump, but generates a complex dynamics, associated to periodic oscillation of a large vortex flow localized in the active region of the pump. Finally, we demonstrate  the existence of an upper bound on the efficiency of such electromagnetic pump, which can never exceed 50\%.
\end{abstract}

\maketitle

\section{\label{sec:intro}Introduction}
Magnetohydrodynamical (MHD) flows are one of the most well known example of energy conversion. In presence of an electrically conducting fluid, it is possible to generate magnetic energy from the kinetic energy of the fluid, or alternatively, to put in motion the fluid by the action of an electromagnetic force \cite{Davidson2001}. Electromagnetic pumping of liquid metals is a good example: By applying an external magnetic field and an electrical current perpendicular to the field, the corresponding Lorentz force $F=J\times B$ will drive the flow. 

It is also possible to electromagnetically drive a flow by applying solely a magnetic field, at the condition that the imposed magnetic field varies both in time and space. In this case, electrical currents in the liquid are induced by the variation of the magnetic flux, rather than injected through electrodes, as in conduction pumps. Such induction flows are ubiquitous, found in both industry \cite{Kliman79} and natural systems \cite{Gissinger19}. In this paper, we will focus on the annular electromagnetic linear induction pump (EMP), which is  considered for secondary cooling systems of fast breeder reactors (FBR), mainly because of the ease of maintenance due to the absence of bearings, seals and moving parts. In such EMPs, sometimes called Einstein-Szilard pumps \cite{Einstein1930} the conducting fluid is generally driven in a cylindrical annular channel by means of a traveling magnetic field imposed by external coils localized on the lateral cylinders (see Fig. \ref{fig1}(a)). Because of its advantages over mechanical pumps, EMPS are considered as a possible candidate for the powering of  intermediate Sodium loop of the future ASTRID nuclear  FBR of generation IV.

The appeal for this technology is however tempered by a major problem:  as the induction effects become large enough compared to dissipation, an instability is produced in the flow, which dramatically reduces the developed flow rate: The fluid suddenly bifurcates from a state of quasi-synchronism with the traveling magnetic field towards a relatively stalled regime \cite{Kirillov80,Araseki00,Araseki04}. Although a theoretical explanation has been formulated for idealized situations \cite{Gailitis76,Zikanov2007,Chu98}, the mechanisms by which this stalling instability occurs in the presence of turbulence are still unclear and represent a very active field of research. Recently, it was shown \cite{Gissinger2016,Rodriguez2016} that this behavior is in fact strongly connected to the phenomenon of 'flux expulsion', which is known to occur in magnetohydrodynamics in presence of varying magnetic fields: similarly to the skin effect produced by an alternating electric current flowing in a conductor, alternating magnetic fields can be expelled from a conducting fluid if the dissipative effects in the fluid are sufficiently small (see for instance \cite{Moffatt82}). 

Although some solutions have been proposed to limit the stalling of EMPs, several questions remain unanswered: first, the exact physical mechanisms by which the instability occurs are unclear and there is currently no theoretical framework for the description of such electromagnetically driven flows in the turbulent regime. Although some experimental results \cite{Araseki00,Araseki04,Ota2004} seem to indicate a very complex dynamics of the flow close to the instability, there is currently no understanding of the type of dynamical regimes observed in these systems. Answering these various questions is a crucial step towards an improvement of the efficiency of electromagnetic pumps. In this perspective, there is strong lack for 3D direct numerical simulation. 

During the last years, our knowledge of such EMPs has been significantly enhanced by the research program PEMDYN installed at CEA, which relies on an experimental Sodium pump with a flowrate up to $1500$ m$^3/$h and a pressure up to $2.5$ bar \cite{Goldsteins2015}, and associated numerical simulations\cite{Delannoy2019, Lopez2018}. 
In this work we present axisymmetric and non-axisymmetric simulations of an annular electromagnetic pump based on the PEMDYN configuration.Complementary, we study an alternative configuration of such pumps, adding a forcing equivalent to a pump with a set of  external coils also placed in the inner cylinder of the annular channel. For both configurations, we investigate the non-axisymmetric destabilization of the pump and its efficiency as a function of the control parameters. 

\section{\label{sec:results}Results}

\begin{figure}
\subfloat[]{\includegraphics[width=7cm]{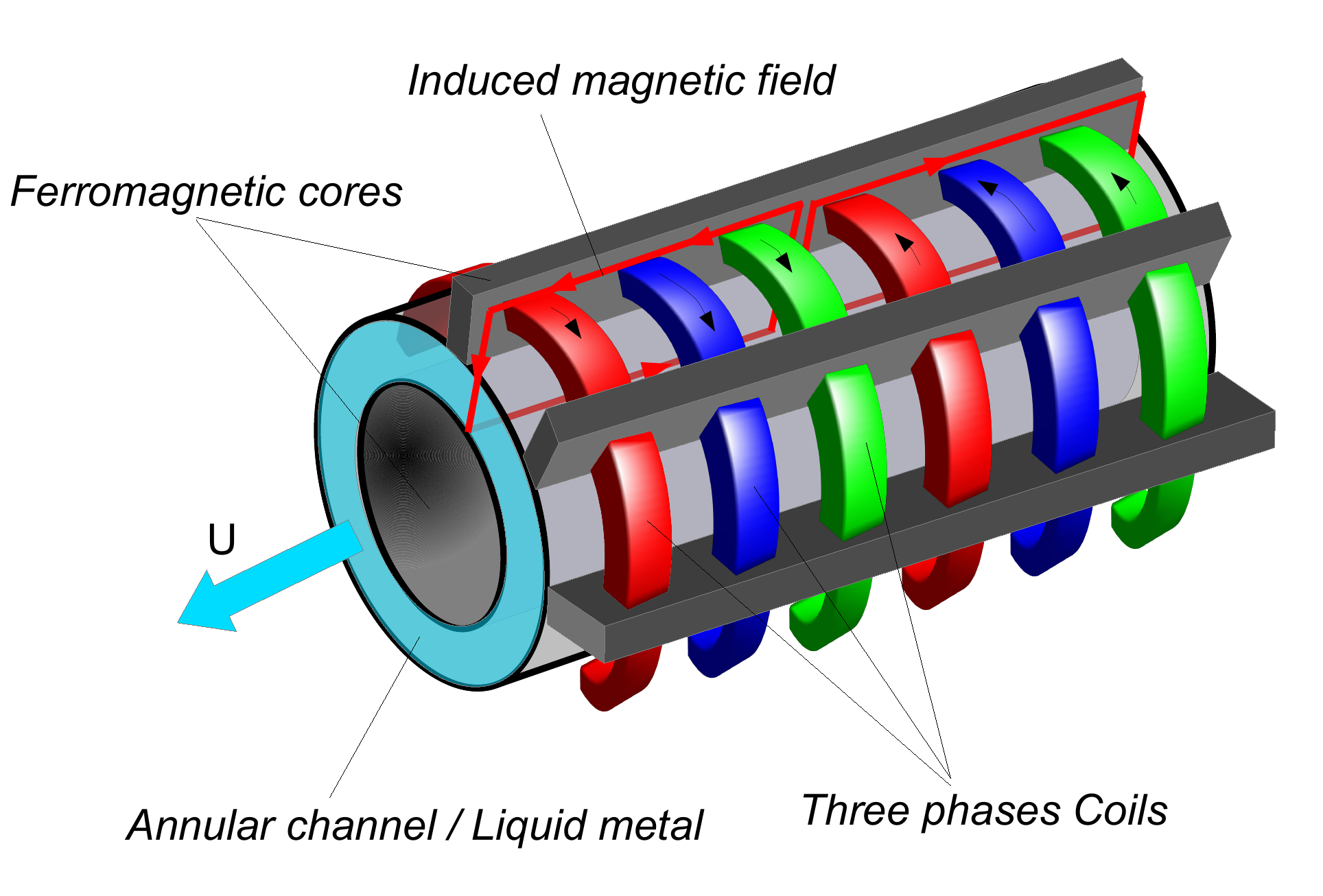}}
  \hspace{0.5cm}
  \subfloat[]{\includegraphics[width=7cm]{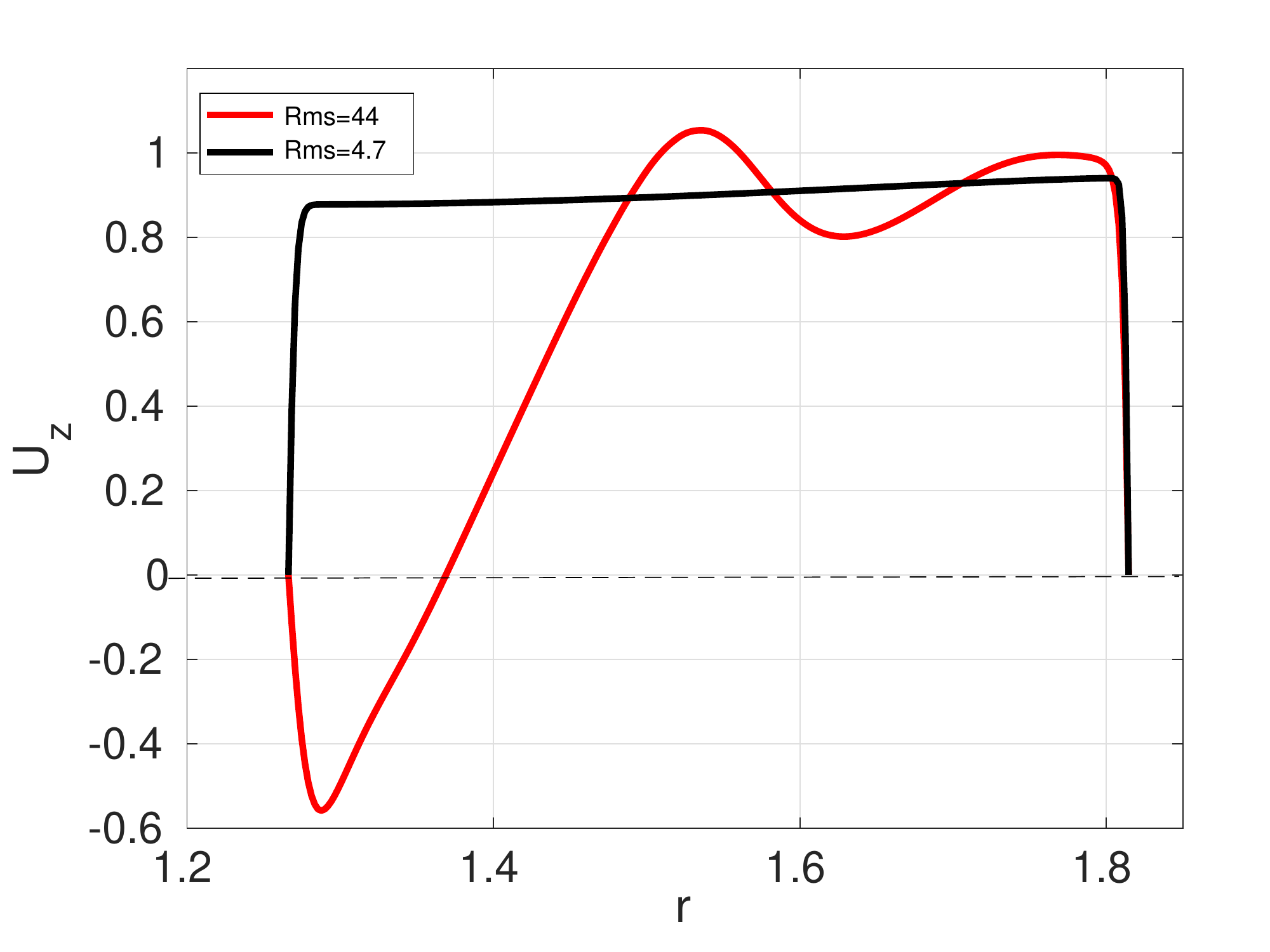}}
  \vspace{0.5cm}
\caption{\label{fig1}(a) Schematic view of a linear induction annular electromagnetic pump. A set of 3-phase coils system located on the external vessel induces a traveling magnetic field which drives the flow confined in the cylindrical annulus in the axial direction. (b) Radial profiles of the time averaged axial velocity $v_z$ for $R_{ms}=4.7$ (black) and $R_{ms}=44$ (red) in axisymmetric simulations with $Ha=1000$, $Re=5000$. Note the stalling of the pump at large $R_{ms}$, associated to flow pumped backwards.}
\end{figure}

A schematic view of a typical annular induction electromagnetic pump is shown in Fig.~\ref{fig1}(a). The liquid metal flows along an annular channel made of two concentric cylinders. A three-phase system of electrical currents placed on the outer cylinder ensures the propagation of a periodic magnetic field traveling in the axial direction. Ferromagnetic cores reinforce the radial component of this induced magnetic field. In order to model such EMP, we consider the flow of an electrically conducting fluid between two concentric cylinders, $r_i$ is the radius of the inner cylinder, $r_o = r_i/\beta$ is the radius of the outer cylinder, and $H$ is the length of the annular channel between the cylinders. In all the simulations reported here, periodic boundary conditions are used in the axial direction. On the cylinders, we consider infinite magnetic permeability boundary conditions, for which the magnetic field is forced to be normal to each boundary. In addition, an azimuthal electrical current $J_\theta=J_0\sin(kz-\omega t)$ is imposed on the outer cylinder (see \cite{Gissinger2016,Rodriguez2016} for more details). Our equations are made dimensionless by a length scale $l_0=r_i(r_o-r_i)$ and a velocity scale $c$, where $c = \omega/k$ is the speed of the traveling magnetic field (TMF). The problem is then governed by the following dimensionless numbers: the magnetic Reynolds number $\epsilon=\mu_0\sigma c l_0$, the Hartmann number, which controls the magnitude of the applied current, defined as $H_a=\mu_0J_0l_0\sqrt{\sigma/\rho\nu}$, and the kinetic Reynolds number $R_e=c l_0/\nu$.
Here, $\mu_0$, $\sigma$ and $\nu$ are respectively the magnetic permeability, electrical conductivity and kinematic viscosity of the fluid. These equations are integrated with the HERACLES code [6], modified to take into account both viscosity and magnetic resistivity \cite{Gissinger11}. Typical resolutions used in the simulations reported in this article span from $(Nr,Nz) = [256,1024]$ for the axisymmetric case to $(Nr,N{\phi},Nz) = [128,256,512]$ in fully 3D simulations.

In a series of recent papers\cite{Gissinger2016,Rodriguez2016}, direct numerical simulations of such EMPS were performed in the axisymmetric case. It was shown that for sufficiently large magnetic and kinetic Reynolds number, the flow looses its homogeneity in the radial direction:  the magnetic field induced by external coils is suddenly expelled from the bulk flow of the pump, and confined to a thin skin depth close to the external cylinder. As a consequence, the fluid close to the outer cylinder stays in synchronism with the TMF, while the fluid close to the inner cylinder exhibits velocity very small, or even opposed to the magnetic driving. It was shown that this state corresponds to a local stalling of the flow located away from the driving coils. This behavior of the pump is illustrated in Fig.~\ref{fig1}(b), which shows the velocity profiles in the synchronous regime (black curve), and after the transition, in the stalled regime (red curve). Note how in the latter case, the fluid close to the inner cylinder is pumped backward compared to the driving TMF. 

\begin{figure}
\centerline{\includegraphics[width=14cm,height=6cm]{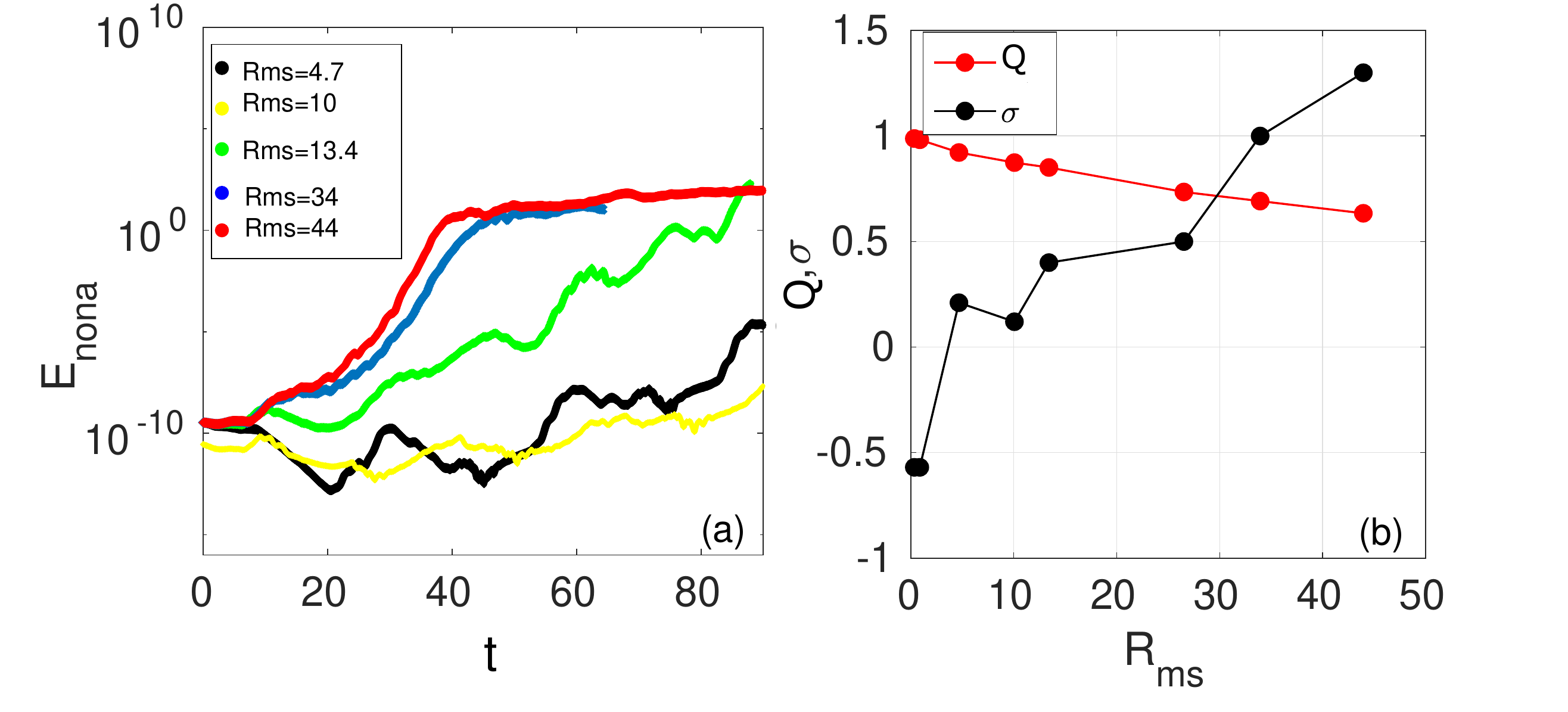}}
\caption{\label{fig2}(a)Total energy of non-axisymmetric modes corresponding to simulations with $\epsilon=60, 80, 90, 110$ and $120$, at fixed $H_a=1000$ and $R_e=5000$ as a function of time, in semi-log scale.(b) Normalized flowrate $Q$ (red curve) for the axial velocity and growth-rate $\sigma$ (black dots) of the total non-axisymmetric energy as a function of $R_{ms}=\epsilon(1-q)$. Note that for $\epsilon$ comprised between $30$ and $120$, $R_{ms}$ is comprised between $0.35$ and $44$.}
\end{figure}

We now turn to the full non-axisymmetric simulations. Fig.~\ref{fig2}(a) shows the total energy contained in the non-axisymmetric modes for  $\epsilon=60, 80, 90, 110$ and $120$, at fixed $H_a=1000$ and $R_e=5000$. Above a critical $\epsilon \approx 90$ the non-axisymmetric modes begin to grow with a growth rate that highly depends on $\epsilon$, as can be observed in fig.~\ref{fig2}(b), which shows both the growth rate of the non-axisymmetric  modes and the corresponding flow-rate as a function of $R_{ms}$, the slip based $R_{ms}$. This number is defined as $R_{ms}=(1-q)\epsilon$, where $q=v/c$ is the slip (\textit{i.e.} the ratio between the velocity $v$ developed by the flow in the annular channel and the velocity of the TMF $c$), and is therefore more representative of the real velocity of the flow. It means that $R_{ms}$ rather than $\epsilon$ should be used when comparing our numerical results to experiments. In  this perspective,  note that the values of $\epsilon$ are quite larger than those found in experimental EMPs, the values of $R_{ms}$ reported in this article are very close to those achieved in such pumps.

For the non-axisymmetric simulations performed, this number is comprised between $0.35$ and $44$. These values where calculated for each run using the mean flow-rate $q$ integrated in a sub-domain of the annular channel which excludes the boundaries both in $r$ and $z$ .

It should be noted that the $R_{ms}$ achieved in the simulations are very close to those obtained in real experimental EMPs, that span from $R_{ms}\sim 1$  in  small laboratory experiments \cite{Pereira19} , to $R_{ms}=11$ for medium size pumps \cite{Goldsteins2015} and $R_{ms}=20$ for large EMPs \cite{Ota2004}. In our simulations a sharp jump of the flowrate is observed for $R_{ms}$ between $12$ and $25$, as clearly shown in fig.~\ref{fig2}(b). The critical value for the appearance of the stalling instability is around $R_{ms}\approx 13$.

\begin{figure}
\centerline{\includegraphics[width=15cm,height=5cm]{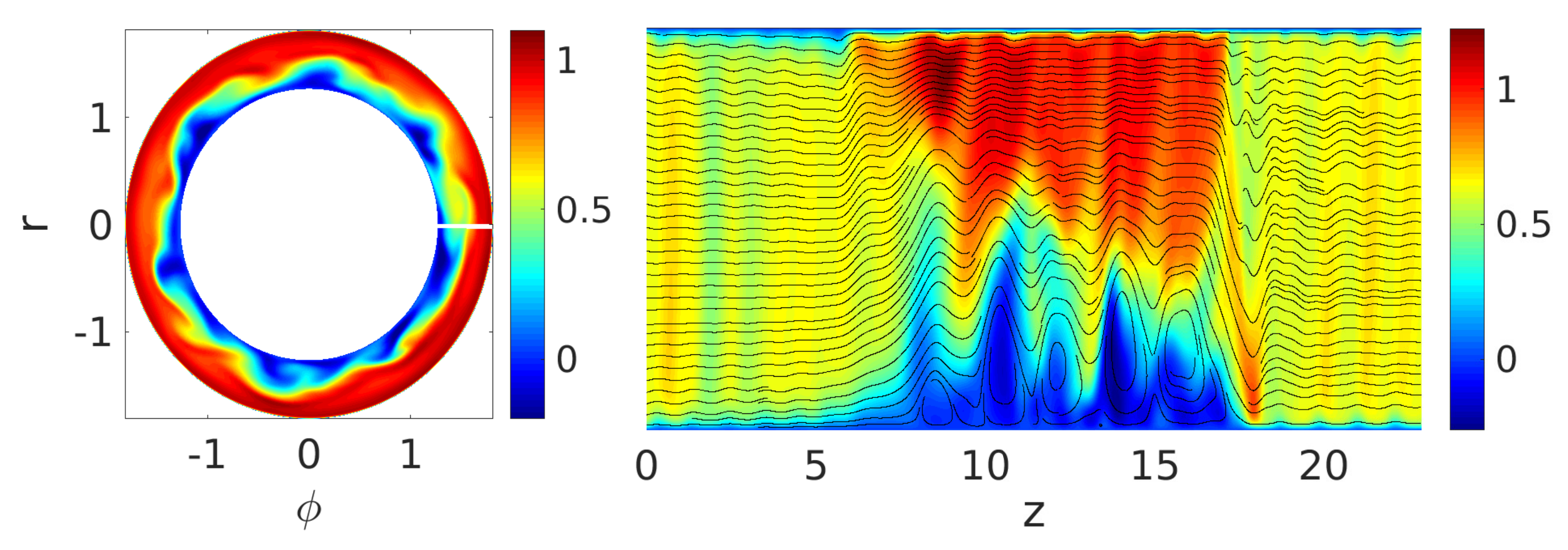}}
\caption{Snapshot of the axial velocity for a simulation with $R_{ms}=44$ ($\epsilon=120$), at fixed $H_a=1000$ and $R_e=5000$ in the $r-\phi$  plane (left) and in the $r-z$ plane (right). Note how the shear generated by the counter flow instability in the axial direction gives rise to vortices that travel in the $\phi$ direction.}
\label{fig3}
\end{figure}

The destabilization of EMPs was previously predicted by various authors as occurring in the azimuthal direction \cite{Gailitis76,Araseki04}, but following a scenario relatively different from the mechanism observed here. In \cite{Araseki04}, it is suggested that the inhomogeneities of the setup along the azimuthal direction (due to the presence of ferromagnetic cores generating non-axisymmetric magnetic field), is responsible for the loss of axisymmetry of the pumped flow. The scenario observed in our DNS is relatively different. First, our setup is completely axisymmetric since  the inlet/outlet conditions, applied magnetic field and applied currents are axisymmetric. In addition, the bifurcation of non-axisymmetric modes always occurs above the critical $\epsilon$ corresponding to the occurrence of the axisymmetric instability described in detail in ref.~\cite{Rodriguez2016}. Fig.~\ref{fig3} shows the axial velocity for a typical simulation at $\epsilon>\epsilon_{c}$, and  illustrates the large scale axisymmetric vortices generated in the flow, leading to an statistically stationary regime in which the flow inside the pump is locally pumped backwards. This local stalling of the flow (i.e. the counter flow observed in the region closer to the inner cylinder) generates a strong shear in  the center of the  $(r-z)$ domain. Non-axisymmetric simulations for different values in the parameter space show that the growth of non-axisymmetric modes immediately starts once the 2D instability is set in, and the base flow takes the form shown in Fig.~\ref{fig3}.

This scenario implies a very different picture than the one known for  typical MHD machines\cite{Gailitis76}, where the destabilization of the flow in the $\phi$ direction is predicted to be a primary bifurcation from the radially homogeneous flow.
Our DNS suggest that the flow instability of turbulent electromagnetic pumps follows a different scenario: the pump first becomes unstable in the radial direction, because local flux expulsion only occurs at the inner boundary not subjected to surface currents (see \cite{Gissinger2016,Rodriguez2016} for more details). This radially inhomogeneous flow then generates a strong shear perpendicular the flow motion. The interface between positive and negative axial velocities becomes unstable and roll over along the azimuthal direction, leading to a strongly non-axisymmetric state involving complex vortices in all three directions. The destabilization of the pump is therefore triggered by a 2D MHD instability that gives rise to a 3D Kelvin-Helmoltz type instability. This transition is sub-critical and highly dependent on the initial conditions. Once the 3D modes have reached their statistically stationary state, they can persist even in the absence of the shear produced by the two-dimensional MHD instability. 

\begin{figure}
\begin{center}
{\includegraphics[width=10cm]{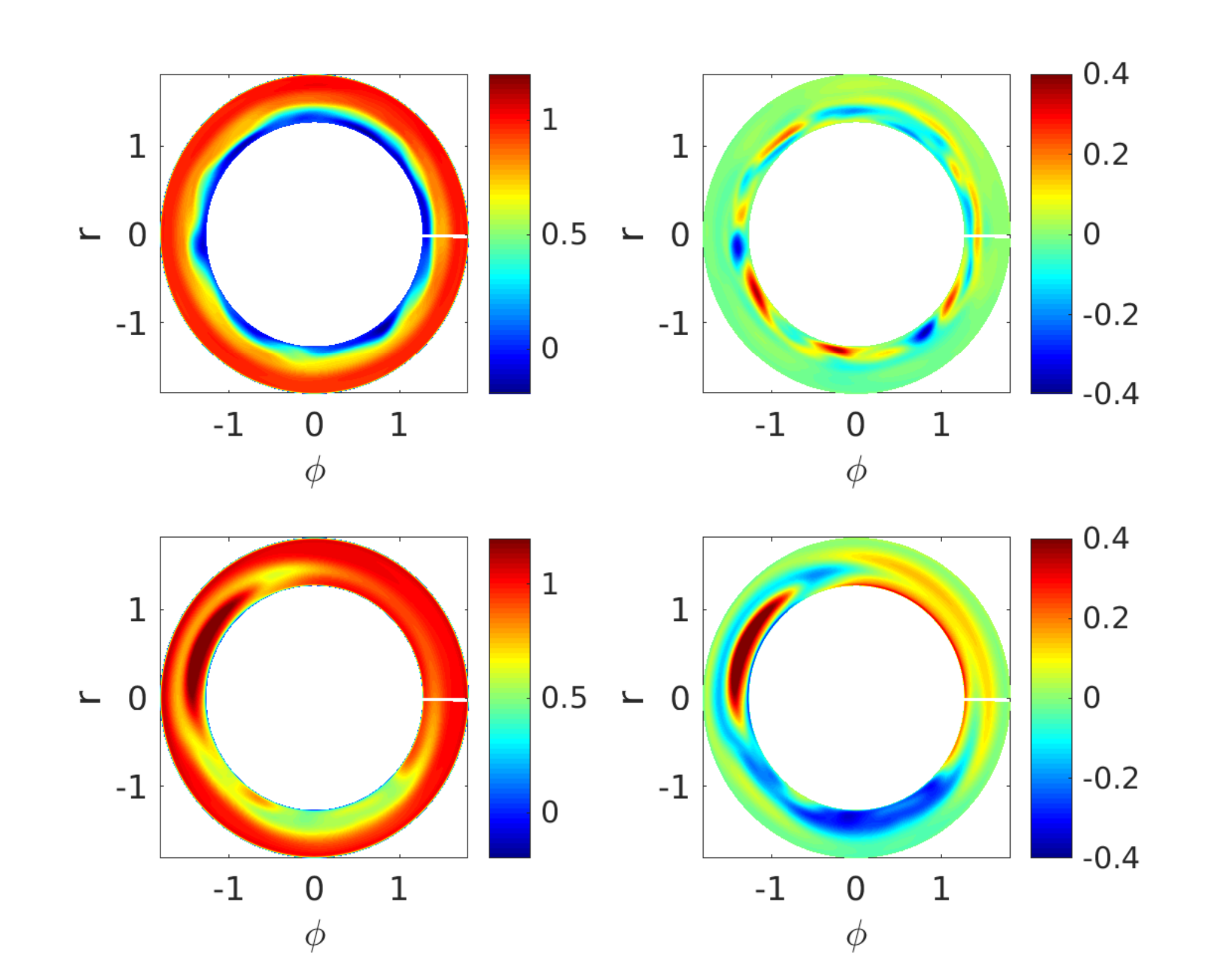}}
\caption{Snapshots of the axial velocity (left) and the nonaxisymmetric velocity (right) in the $(r-\phi)$ plane for $R_{ms}=10$ (top) and $R_{ms}=4.7$ bottom. Stability of the base flow is slowly recovered as we decrease $\epsilon$, whereas nonaxisymmetric modes remain.}
\label{fig4}
\end{center}
\end{figure}

To illustrate this, we performed simulations  decreasing $\epsilon$ at constant $H_a=1000$, and using as initial condition for each run the last stationary state of the previous one (i.e. at larger $\epsilon$). These runs were performed starting at $\epsilon=120$ and ending at $\epsilon=30$ (corresponding to $R_{ms}=44$ and $R_{ms}=0.35$, respectively) using $\epsilon=20$ at the decreasing step. As we move to smaller values of $\epsilon$, it can be seen that the non-axisymmetric energy remains constant for values outside the MHD instability pocket, achieving up to $40 \%$ of the total energy but following a slow recover to synchronism of the base flow. In fig.~\ref{fig4} we show two snapshots of the total and non-axisymmetric axial velocity for runs with $R_{ms}=10$ and $R_{ms}=4.7$ ($\epsilon=80$ and $\epsilon=60$, respectively). For early times of the run with $R_{ms}=10$ a return flow is still present in the axial direction, which favor the development of azimuthal vortices. As we go to $R_{ms}=4.7$ the base flow is nearly homogeneous, but non-axisymmetric modes are still present.

\subsection{Double sided configuration}

These results suggest that the shear produced by the flux expulsion instability is the key ingredient for the loss of synchronism in electromagnetic pumps. In order to suppress this instability and increase the efficiency of EMPs, a natural choice is to enhance the magnetic field applied by the external coils. The latter can be achieved by adding external coils along the inner cylinder, in addition to the outer coils. We will refer to this type of forcing as the double sided (ds) configuration, by opposition to the  single sided (ss) configuration studied above. Numerically, $ds$ configuration is modeled by imposing on the inner cylinder $r=r_i$ a surface current $J_s=J_0\sin(kz-\omega t)$ of magnitude similar to the one applied at $r=r_o$.

\begin{figure}
\begin{center}
\includegraphics[width=14cm,height=6cm]{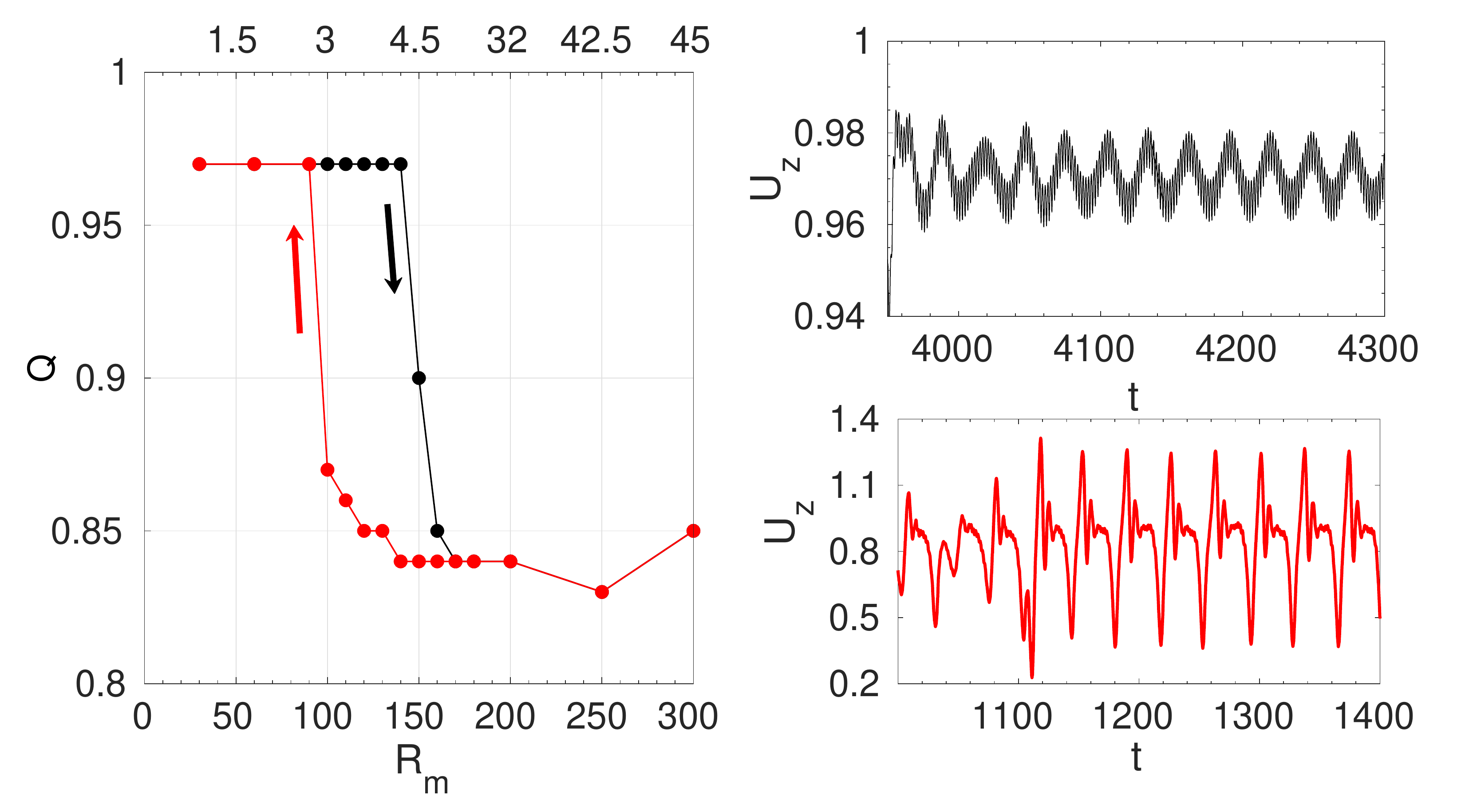}\\
(a)\hspace{3.5cm}(b)
\caption{(a) Normalized flowrate $Q$ for the mean axial velocity in the center of the pump for runs using the $ds$ configuration. The higher branch, denoted by the black curve indicates runs with increasing $\epsilon$, whereas the low branch is represented by the red curve, corresponding to decreasing $\epsilon$. Each point of the curves represents a concatenated run in the parameter space, varying $\epsilon$ between $30$ and $300$ at fixed $H_a=1000$ and $R_e=5000$. The corresponding $R_{ms}$ for these runs, comprised between $R_{ms}=0.47$ and $R_{ms}=45$ for both branches, is indicated by the (forced) ticks on the upper $x$ axys, for the upper branch. (b) Time series of the normalized axial velocity corresponding to points with $\epsilon=150$ for the higher branch (top) and the lower branch (bottom).}
\label{fig5}
\end{center}
\end{figure}

The parameter space explored for the $ds$ configuration is similar to the one explored for the $ss$ configuration shown in fig.~$5$ of ref.~\cite{Rodriguez2016}. For fixed $H_a=1000$, it comprehends values of $\epsilon=30,60$ and $90$,  increasing up to $\epsilon=300$ with a step of $\epsilon=10$ for $\epsilon \geq 90$. Also, runs using fixed $\epsilon=90$ and $120$, where performed, 
varying the Hartmann number from $50$ to $800$. 
In total, $29$ axisymmetric runs where performed for the $ds$ configuration, all with fixed $R_e=5000$.

Similarly to the single sided configuration shown in ~\cite{Rodriguez2016}, flows with very small values of the velocity in the whole channel are observed at very low Hartmann number. At sufficiently large $H_a$, a sharp transition from homogenous to inhomogeneous flows can be triggered as $\epsilon$ is increased, which leads to a low frequency (LF) pulsation of the flow. 
This sharp transition is depicted in Fig.~\ref{fig5}(a), which shows a bifurcation diagram for the mean axial flowrate in the center of the channel, as we increase (black curve) and decrease (red curve)  $\epsilon$ at fixed $H_a=1000$. The transition in $\epsilon$ is again subcritical, dropping from nearly synchronous velocities to considerably smaller values of the flow rate at a critical $\epsilon\sim 160$ for the upper branch and $\epsilon\sim 90$ for the lower branch. Slip based magnetic Reynolds number calculated for runs in the upper branch is denoted in the superior axes of fig.~\ref{fig5}(a), ranging from $R_{ms}= 0.47$ to $R_{ms}=45$. For runs in the lower branch, $R_{ms}$ is comprised between the same values, taking values between $1.1\leq R_{ms} \leq 22$ for 
$60 \leq \epsilon \leq 160$, where the bifurcation occurs.  

\begin{figure}
\centerline{\includegraphics[width=15cm,height=5cm]{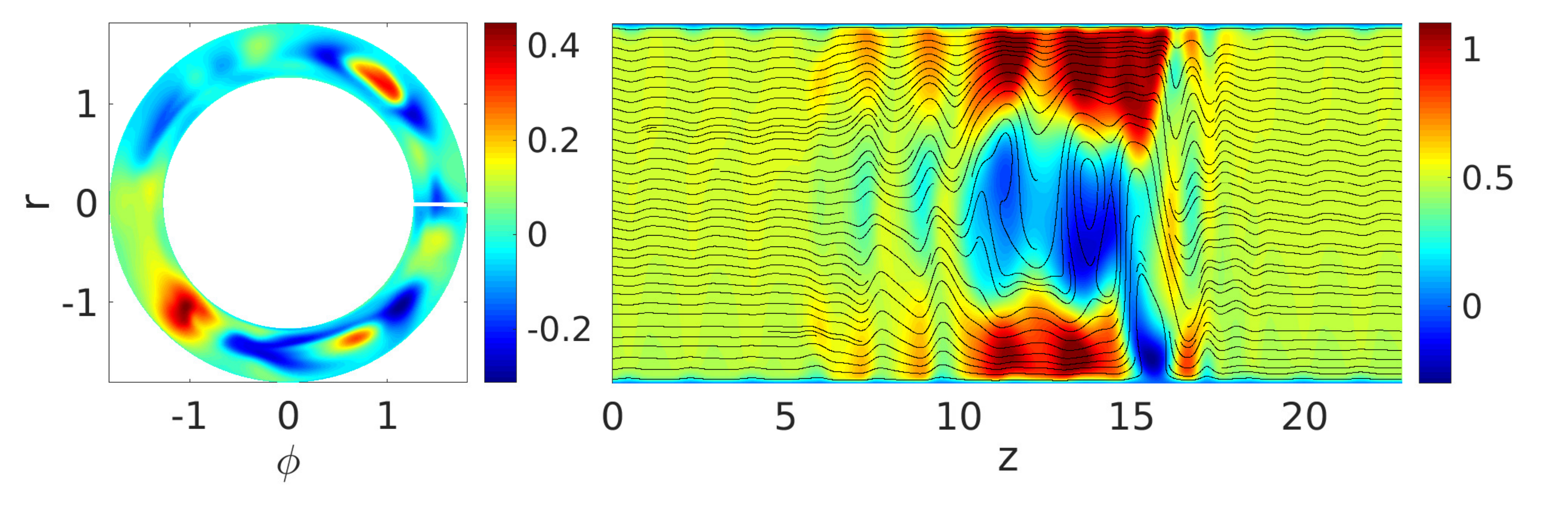}}
\caption{Snapshots of the nonaxisymmetric (left) and axial velocity (right) in the $(r-\phi)$ and the $(r-z)$ plane respectively, for the $ds$ configuration at $\epsilon=120$, $H_a=1000$ and $R_e=5000$.}
\label{fig6}
\end{figure}

Figure~\ref{fig5}(b) shows the time series of the normalized velocity in $z/2$, averaged in the radial direction for the homogeneous (top) and inhomogeneous (bottom) LF state. The pulsation of the axial velocity, characterized by typical frequencies nearly $10$ times smaller than the forcing frequency, is characterized by a vortex appearing and disappearing in the central region of the pump, and leads to a negative flow in the center of the pump. This is illustrated by Fig.~\ref{fig6}, which shows the snapshots of the nonaxisymmetric and the axial velocity for a simulation with $\epsilon=120$ (corresponding to an $R_{ms}=44$), $H_a=1000$, and $R_e=5000$. 
Note that a similar LF pulsation is also observed for the $ss$ simulations, but only  in the vicinity of the critical $\epsilon_c$ for the stalling instability described above.

Interestingly, the periodic pulsation of the vortex traveling in the $z$ direction also produces a spontaneous counter flow in the center of the pump, that generates the same vortices  in the $r-\phi$ plane (see left panel of Fig.~\ref{fig6}).
Therefore, our results indicate that the $ds$ configuration is not prone to suppress the 3D destabilization of the flow. It should be noted however that the magnitude of the non-axisymmetric component of the flow is much smaller than the one observed in single-sided pumps, leading to larger time-averaged flow rates.  

\begin{figure}
\centerline{\includegraphics[width=12cm]{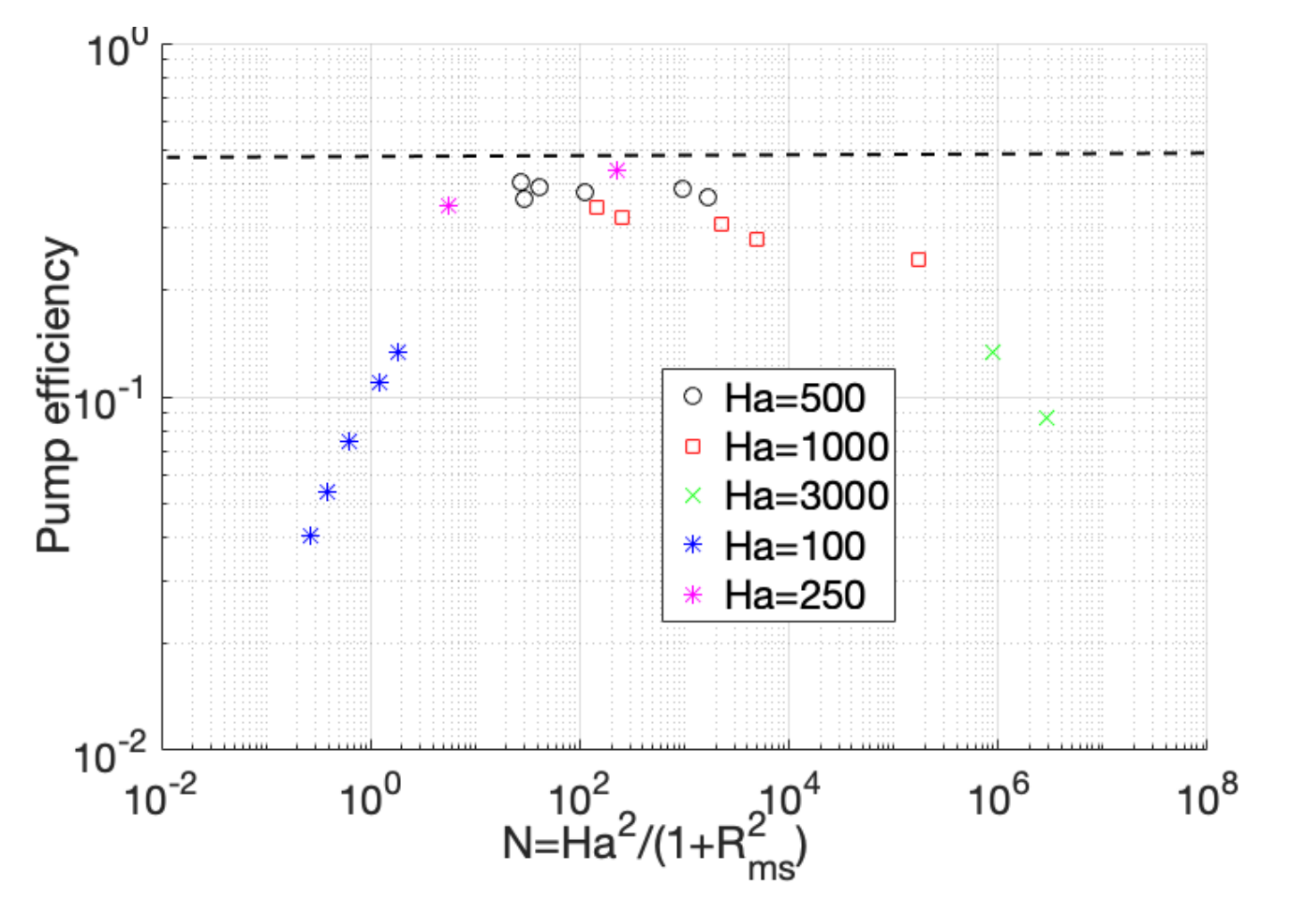}}
\caption{Pump efficiency as a function of the dimensionless parameter $N$. Note that this efficiency is bounded by the maximum value of $50\%$.}
\label{efficiency}
\end{figure}

Finally, the performance of the pump can be estimated by computing the efficiency $\eta=L/P$, where $L = \langle\mathbf{u\cdot}[(\mathbf{\nabla\times b)\times
    b}]/\mu_0\rangle_V$ is the power of the Lorentz force, and $P=\frac{\eta}{\mu_0}\langle\mathbf{b\times}\mathbf{(\nabla\times b)}\rangle_S$ is the injected power due to surface currents imposed at the cylinders.
$\langle\rangle_V$ denotes spatial average over the annulus (in the active region of the pump), while $\langle\rangle_S$ denotes spatial average over the surface of inner and outer cylinders. The efficiency is deeply linked  to the PQ characteristic of the pump, given by \cite{Gailitis76} :

\begin{equation}
\frac{Ha^2}{1+R_{ms}^2}(1-q)-\gamma q^2=0
\end{equation}

where $\gamma$ is a free parameter related to the Reynolds number and the drag coefficient $C_D$ in the channel, and $q=u_0/c$ is the normalized bulk velocity. Similarly to what was recently observed in a slightly different setup \cite{Sandeep2018}, Fig.\ref{efficiency}  shows that there is a fairly good rescaling of all datas when the efficiency of the pump is plotted as a function of the dimensionless number $N=Ha^2/(1+R_{ms}^2)$ suggested by the PQ characteristic above. In addition, $\eta$ exhibits a non-monotonic behavior as a function of $N$, with a maximum value close to $1/2$. The existence of such a bound can be derived theoretically (see \cite{Sandeep2018}), but it can also be understood with simple arguments: for sufficiently large Hartmann number, the squared shape of the magnetized profile implies that most of the viscous dissipation $D_\nu=\rho\nu \int_V(\nabla\times u)^2$ occurs in the Hartmann boundary layer of thickness $\delta$, such that $D_\nu\sim\rho\nu u_0^2/\delta$. Similarly, the ohmic dissipation in the boundary layer is given by $D_\eta\sim \sigma B_0^2c^2\delta$ such that the $D_\nu\sim D_\eta$ (in the layer). In the bulk flow, exact synchronism is never achieved, such that there is always an extra-contribution to the ohmic dissipation due to the induced currents $j\sim \sigma(c-u_0)B_0$. The ratio of viscous to ohmic dissipation is thus unity at best,  leading to an upper bound of $50\%$ on the pump efficiency. It is interesting to note that this theoretical bound seems to be observed in many experimental pumps \cite{Ota2004,Fanning2003}.

\section{\label{sec:conclu}Conclusions}

In this article we present the results of non-axisymmetric direct numerical simulations of an MHD flow driven by a travelling magnetic field in an annular channel, extending the results previously presented in ~\cite{Gissinger2016,Rodriguez2016} for axisymmetric laminar and turbulent flows to fully 3D simulations. 

For the non-axisymmetric turbulent case studied here, we have shown that above a critical $\epsilon$ corresponding to the occurrence of the axisymmetric stalling instability described in ref.~\cite{Rodriguez2016}, there is a  bifurcation of non-axisymmetric modes characterized by the development of large scale axisymmetric vortices in all three directions. Thus, the destabilization of the pump is triggered by a 2D MHD instability that gives rise to a 3D Kelvin-Helmoltz type instability. We have also found that this transition is sub-critical and highly dependent on the initial conditions. Once the 3D modes have reached their statistically stationary state, they can persist even in the absence of the shear produced by the two-dimensional MHD instability. This novel results imply a very different picture than the one known for typical MHD machines\cite{Gailitis76}, where the destabilization of the flow in the $\phi$ direction is predicted to be a primary bifurcation from the radially homogeneous flow. 

In order to suppress the instability and increase the efficiency of EMPs, we propose an alternative configuration for the EM forcing, adding external coils along the
inner cylinder. For this, we performed non-axisymmetric DNS using the same forcing at both radial ends of the annular channel. 
Our results show that above the same $\epsilon_c$, a low frequency pulsation appears, characterized by a pulsating vortex traveling in the $z$ direction. Even  though no stationary counter flow is generated, the pulsating vortex leads to negative axial velocities in the center of the pump. As a result of this spontaneous counter flow, vortices in the $r-\phi$ direction appear. Therefore, our results suggest that the use of coils at both the inner and outer cylinders of the pump does not suppress the 3D instability. However, the magnitude of the non-axisymmetric component of the flow is much smaller than the one observed in single-sided pumps, leading to larger time-averaged flow rates.

Finally we showed that a full rescaling of our data is obtained when the efficiency of the pump is  plotted as a function of the dimensionless number $N=H_a^{2}/1+R_{ms}^2$. Surprisingly, it is  found that this efficiency can never exceed $50 \%$, confirming a recent theoretical prediction and showing a very good agreement with previous experimental results. Since our numerical results reported here indicate secondary bifurcations towards non-axisymmetric states in both single-sided or double-sided configurations for EMPs, the consideration of these results could be useful to improve real pumps efficiency. 

\begin{acknowledgments}
This  work  was  supported  by  funding  from  the  French  program  Retour  Postdoc managed  by  Agence  Nationale  de  la
Recherche (Grant ANR-398031/1B1INP), and the DTN/STPA/LCIT of Cea Cadarache.  The present work benefited from the
computational support of the HPC resources of GENCI-TGCC-CURIE (Project No.  t20162a7164) and MesoPSL financed by
the Region Ile de France where the present numerical simulations have been performed 
\end{acknowledgments}



\end{document}